\documentclass[amsmath,amssymb,prd,a4paper,onecolumn,preprint]{revtex4}

\usepackage{graphicx}
\usepackage{amsmath}
\usepackage{epstopdf}
\usepackage{amsfonts,amssymb}
\usepackage[utf8]{inputenc}
\usepackage{pstricks}
\usepackage{color}

\newcommand{\be}{\begin{equation}}
\newcommand{\ee}{\end{equation}}
\newcommand{\ben}{\begin{eqnarray}}
\newcommand{\een}{\end{eqnarray}}

\def\MeV{\mbox{ MeV}} 
\def\GeV{\mbox{ GeV}} 
\def\mb{\mbox{ mb}} 
\newcommand{\pslash}{\not{\hbox{\kern-2.3pt $p$}}}
\newcommand{\pdslash}{\not{\hbox{\kern-2pt $\partial$}}}

\begin{document}


\title{$X_J(2900)$ states in a hot hadronic medium}

\author{Luciano M. Abreu$^{a}$}

\affiliation{$^{a}$Instituto de F\'{i}sica, Universidade Federal da Bahia, Campus Ondina, 40170-115, Salvador, Bahia, Brazil}

\begin{abstract}

%
%
%
%
%
%
%
%
%
%
%

In this work we investigate the hadronic effects on the $X_{J=0,1} (2900)$ states in heavy-ion collisions. We make use of Effective Lagrangians to estimate the cross sections and their thermal averages of the processes $X_J \pi \to \bar{D}^{*} K , K^{*} \bar{D}  $,  as well as those of the corresponding inverse processes, considering also the possibility of different isospin assignments ($I=0,1$). We complete the analysis by solving the rate equation to follow the time evolution of the  $X_J (2900)$  multiplicities  and determine how they are affected by the considered reactions during the expansion of the hadronic matter. We also perform a comparison of the $X_J (2900)$ abundances considering them as hadronic molecular states ($J=0$ as a $S$-wave and $J=1$ as a $P$-wave) and tetraquark states at kinetic freeze-out.

\end{abstract}

\maketitle


\section{Introduction}

\label{Introduction}


Very recently, The LHCb collaboration has reported the observation of an exotic peak in the $D^- K^+ $ invariant mass spectrum of the $B^+ \to D^+ D^- K^+ $ decay with statistical significance of more than $5\sigma$~\cite{Aaij:2020hon,Aaij:2020ypa}. By using an amplitude model based on Breit-Wigner formalism, this peak has been fitted to two resonances denoted as $X_{0,1}(2900)$, with the corresponding quantum numbers, masses, and widths:
\ben
J^P = 0^+ : \,\, M = (2866\pm 7) \MeV, \,\, \Gamma = (57\pm 13) \MeV ; \nonumber \\
J^P = 1^- : \,\, M = (2904\pm 5) \MeV, \,\, \Gamma = (110 \pm 12) \MeV  .   \nonumber
\label{Xstates}
\een
Their minimum valence quark contents are of four different flavors, i.e. $\bar{c}\bar{s} u d $, giving them the unequivocal statuses of the first observed unconventional hadrons which are fully open charm tetraquarks.

Since the experimental discovery of these states, a debate has opened concerning their internal structure. Several studies have been performed examining distinct interpretations and employing different theoretical approaches. For instance, 
Refs.~\cite{Karliner:2020vsi,Zhang:2020oze,He:2020jna,Chen:2020aos,Wang:2020xyc,Burns:2020xne,Mutuk:2020igv} interpreted the  $X_{0}(2900)$ and/or $X_{1}(2900)$  as compact tetraquarks resulting from the binding of a diquark and an antidiquark, in either fundamental, radially or orbitally excited configurations. On the other hand, due to the proximity to $\bar{D}^* K^*$ threshold, a natural interpretation proposed by other authors was to suppose that they are bound states of spin-1 charmed and $ K^*$ mesons~\cite{Chen:2020aos,Liu:2020nil,Huang:2020ptc,He:2020btl,Hu:2020mxp,Molina:2020hde,Xue:2020vtq,Agaev:2020nrc,Xiao:2020ltm}. In particular, the $X_{0}(2900)$ was explained as a $S$-wave $\bar{D}^* K^*$ hadronic molecule, while the  $X_{1}(2900)$  was described as a $\bar{D}_1 K$ \cite{He:2020btl} or a $P$-wave $\bar{D}^* K^*$ \cite{Huang:2020ptc} hadronic molecule. Besides, Refs.~\cite{Liu:2020orv,Burns:2020epm} suggested that kinematic effects caused by triangle singularities might be the mechanism responsible for the production of this exotic peak detected by LHCb Collaboration.

To determine the structure of the $X_{0,1}(2900)$ states (meson molecules, 
compact tetraquarks, due to kinematic effects or a mixture) more experimental information is required, as well as more theoretical studies to estimate mensurable quantities associated to certain properties that discriminate these different interpretations. 
In this sense, as pointed in Ref. \cite{Cho:2017dcy} (see also \cite{CMS:2019vma}), heavy-ion collisions (HICs) appear as a promising environment to investigate the hadronic production of exotic states. After the phase transition from nuclear matter to a locally thermalized state of deconfined quarks and gluons, the so-called quark-gluon plasma (QGP), heavy quarks coalesce to form bound states and possibily exotic states at the end of the QGP phase. After that, the multiquark states interact with other hadrons during the hadronic phase, and can be destroyed in collisions with the comoving light mesons, but can also be produced through the inverse processes~
\cite{Chen:2007zp,ChoLee1,XProd1,XProd2,Abreu:2017cof,Abreu:2018mnc}.  So, their final multiplicities depend on the interaction cross sections, which in principle depend on the spatial configuration of the quarks. Thus, the evaluation of the interactions of $X_{0,1}(2900)$ states with hadron matter and the measurement  of their 
abundances might be useful to  determine the structure of these states. For example, Refs.~\cite{ChoLee1,XProd2} suggested that the $X(3872)$ multiplicity  at the end of the QGP phase is reduced due to the interactions with the
hadron gas, and  if it was observed in HICs, it would be most likely a molecular state, since this structure would be dominantly produced at the end of hadron phase by hadron coalescence mechanism.

Motivated by the discussion above, in the present work we intend to analyze the hadronic effects on the $X_J (2900)$ states in HICs. We make use of Effective Lagrangians to calculate the cross sections and their thermal averages of the processes $X_J \pi \to \bar{D}^{*} K , K^{*} \bar{D}  $,  as well as those of the corresponding inverse processes. Considering also the possibility of different isospin assignments ($I=0,1$), we analyze dependence of the magnitude of the cross sections and their thermal averages with the quantum numbers. We then solve the rate equation to follow the time evolution of the  $X_J (2900)$  multiplicities  and determine how they are affected by the considered reactions during the expansion of the hadronic matter. We also perform a comparison of the $X_J (2900)$ abundances considering them as hadronic molecular states ($J=0$ as a $S$-wave and $J=1$ as a $P$-wave) and tetraquark states at kinetic freeze-out.

The paper is organized as follows. In Section~\ref{Hadronic Effects} we describe the effective formalism and calculate the cross sections of  $X_J (2900) - \pi$ absorption and production and their thermal averages. With these results, in  Section~\ref{abundance}  we solve the rate equation and follow the time evolution of the $X_J (2900)$ abundances.  Finally, in  Section~\ref{Conclusions} we present some  concluding remarks.

\section{Hadronic effects on the  $X_J(2900)$}
\label{Hadronic Effects}

\subsection{The effective formalism}
\label{Effective Formalism}

\begin{figure}[!ht]
    \centering
        \includegraphics[width=6.0cm]{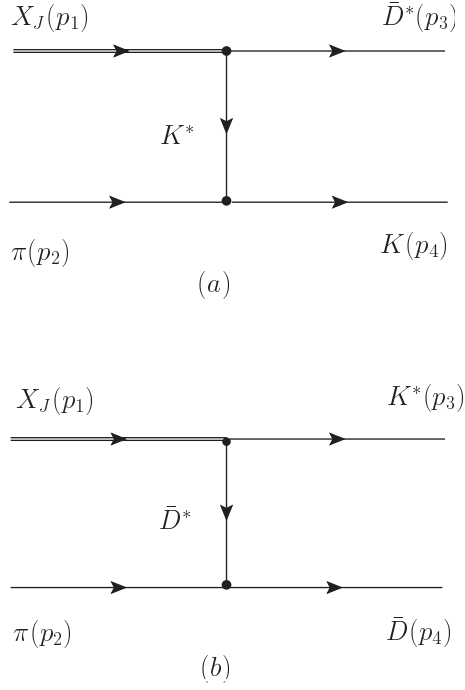}
        \caption{Diagrams contributing to the process $X_J \pi \to \bar{D}^{\ast} K $ [diagram (a)] and $X_J \pi \to K^{\ast}  \bar{D} $ [diagram (b)], without specification of the charges of the particles.}
\label{DIAG1}
\end{figure}

We start by analyzing the interactions of $X_J(2900)$ with the surrounding hadronic medium composed of the lightest pseudoscalar meson ($\pi$). We expect that the reactions involving pions provide the main contributions to the $X_J$ in hadronic matter, due to their large multiplicity with respect to other light hadrons~\cite{Chen:2007zp,ChoLee1,XProd1,XProd2,Abreu:2017cof,Abreu:2018mnc}. 
In particular, we focus on the reactions $X_J \pi \to \bar{D}^{\ast} K $ and $X_J \pi \to K^{\ast}  \bar{D} $, as well as the inverse processes. In Fig.~\ref{DIAG1} we show the lowest-order Born diagrams contributing to each process, without the specification of the particle charges. To calculate the respective cross sections, we make use of the effective hadron Lagrangian framework. Accordingly, we follow Refs.~\cite{Chen:2007zp,Huang:2020ptc} and employ the effective Lagrangians involving $\pi$, $K$, $D$, $K^*$  and $D^*$ mesons,
\begin{eqnarray}\label{Lagr1}
{\mathcal{L}}_{\pi D D^* } &=& i g_{\pi D D^*} D_\mu ^* \vec{\tau} \cdot \left(  \bar{D} \partial^\mu
\vec{\pi} -   \partial^\mu \bar{D} \vec{\pi} \right) + H. c., 
\nonumber\\
{\mathcal{L}}_{\pi K K^* } &=& i g_{\pi K K^*}  \bar{K}_\mu ^* \vec{\tau} \cdot \left(  K \partial^\mu
\vec{\pi} -   \partial^\mu K \vec{\pi} \right) + H. c., 
\end{eqnarray}
where $\vec{\tau}$ are the Pauli matrices in the isospin space; $\vec{\pi}$ denotes the pion isospin triplet; and $D^{(\ast)} = (D^{(\ast) 0}, D^{(\ast) +}  ) $ and $K^{(\ast)} = (K^{(\ast) +}, D^{(\ast) 0}  )^T $ represent the isospin doublets for the pseudoscalar (vector) $D^{(\ast) }$ and $K^{(\ast) }$ mesons, respectively. The coupling constants are determined from decay widths of $D^{\ast }$ and $K^{\ast }$, having the following values~\cite{Chen:2007zp,ChoLee1}:  $g_{\pi D D^*} = 6.3$ and $g_{\pi K K^*} = 3.25$.


The couplings including the $X_J(2900)$ states have been built in order to yield the transition matrix elements $X_J \rightarrow \bar{D}^{\ast 0} K^{\ast 0} , D^{\ast -} K^{\ast +} $~\cite{Huang:2020ptc},   
\begin{eqnarray}\label{Lagr2}
{\mathcal{L}}_{X_0 \bar{D}^{\ast} K^{\ast} } &=& i g_{X_0 \bar{D}^{\ast} K^{\ast}} C_Y ^{(I)} X_0 \bar{D}_\mu ^*  K^{\mu \ast} + H. c., 
\nonumber\\
{\mathcal{L}}_{X_1 \bar{D}^{\ast} K^{\ast} } &=& i g_{X_1 \bar{D}^{\ast} K^{\ast}} C_Y ^{(I)} X_1 ^{\nu} \bar{D}_\mu ^*  \overleftrightarrow{\partial_{\nu}}  K^{\mu \ast}.
\end{eqnarray}
In the expressions above, $X_0$ and $X_1$ denote respectively the $X_0(2866)$ and $X_1(2904)$ states; this notation notation will be used henceforth. Since there is no experimental information yet available on their isospin $(I)$, we consider here the different isospin assignments: $I=0,1$. In this sense, $C_Y ^{(I)}$ labels the isospin factor according to the value of $I$ and the respective channel $Y= \bar{D}^{\ast 0} K^{\ast 0}, D^{\ast- } K^{\ast +}$. 
The values of the coupling constants are chosen according to the analysis based on the compositeness condition performed in Ref.~\cite{Huang:2020ptc}, in which for a cutoff $\alpha = (1.0 \pm 0.1) \GeV$ we have $g_{X_0 \bar{D}^{\ast} K^{\ast}} = 3.82 \substack{-0.16 \\ +0.11} \GeV$ and $g_{X_1 \bar{D}^{\ast} K^{\ast}} = 7.84 \substack{-0.30 \\ +1.00} \GeV$.

With the effective Lagrangians introduced above, the amplitudes of the processes shown in Fig.~\ref{DIAG1} can be calculated; the are given by
\begin{eqnarray}
   \mathcal{M}_a & = &  \tau _{rs} ^{(i)} C_Y ^{(I)} g_{X_J \bar{D}^{\ast} K^{\ast}} g_{\pi K K^*} (p_2 + p_4)^{\mu} \nonumber \\
   & & \times \frac{1}{t - m_{K^*} ^2 + i m_{K^*} \Gamma_{K^*} }   \nonumber \\
   & & \times \left( -g_{\mu \nu} + \frac{(p_1-p_3)_{\mu }(p_1-p_3)_{\nu } }{m_{K^*} ^2} \right)    \epsilon_{ \mu} (p_3) {\mathcal{A}}^{(J)} , \nonumber \\
   \mathcal{M}_b & = &  \tau _{rs} ^{(i)} C_Y ^{(I)} g_{X_J \bar{D}^{\ast} K^{\ast}} g_{\pi D D^*} (p_2 + p_4)^{\mu} \nonumber \\
   & & \times \frac{1}{t - m_{\bar{D}^*} ^2 + i m_{\bar{D}^*} \Gamma_{K^*} }   \nonumber \\
   & & \times \left( -g_{\mu \nu} + \frac{(p_1-p_3)_{\mu }(p_1-p_3)_{\nu } }{m_{\bar{D}^*} ^2} \right)    \epsilon_{ \mu} (p_3) {\mathcal{A}}^{(J)} , \nonumber \\
      \label{ampl}
 \end{eqnarray}
where $\tau _{rs} ^{(i)}$ is the isospin factor related to $i$-th component of $\pi$ isospin triplet and $r(s)$-th component of $D^{(\ast) }$ and $K^{(\ast) }$ isospin doublets; $p_1$ and $p_2$ are the momenta of initial state particles, 
while $p_3$ and $p_4$ are those of final state particles; $t$ is one of the Mandelstam variables: $s = (p_1 +p_2)^2, t = (p_1 - p_3)^2,$ and $u = (p_1-p_4)^2$; and $ \mathcal{A}^{(J)}=  \{ 1, -\epsilon (p_1) \cdot (q - p_3) \}$ for $X_0$ and $X_1$, respectively. 

We determine the isospin coefficients $\tau _{rs} ^{(i)}$ and $ C_Y ^{(I)}$ of the processes reported in Eq.~(\ref{ampl}) by considering the charges $Q_{1f}$ and $Q_{2f}$ for each of the two particles  in final state~\cite{XProd1}, whose combination must yield the total charge $Q = Q_{1f} + Q_{2f} = 0, +1, -1$.  Thus, there are four possible charge configurations $(Q_{1f}, Q_{2f})$ for each reaction in Eq.~(\ref{ampl}). In Table~\ref{Tab1} are listed the values of $\tau _{rs} ^{(i)}$ and $ C_Y ^{(I)}$ for the possible configurations of each process.

\begin{table}
\caption{Isospin coefficients $\tau _{rs} ^{(i)}$ and $ C_Y ^{(I)}$ of the processes described in Eq.~(\ref{ampl}) by considering the charges $Q_{1f}$ and $Q_{2f}$ for each of the two particles in final state~\cite{XProd1}.}\label{Tab1}
\begin{tabular}{c|c|c|c|c}
Process & $(Q_{1f}, Q_{2f})$  & $\tau _{rs} ^{(i)}$ & $ C_Y ^{(I=0)}$  & $ C_Y ^{(I=1)}$  \\
\hline
\hline
a & $(0,+)$ & $-1$ & $ \frac{1}{\sqrt{2}}$ & $ - \frac{1}{\sqrt{2}}$ \\
  & $(0,0)$ & $\frac{1}{\sqrt{2}}$ & $ \frac{1}{\sqrt{2}}$ & $ - \frac{1}{\sqrt{2}}$ \\
  & $(-,+)$ & $-\frac{1}{\sqrt{2}}$ & $ \frac{1}{\sqrt{2}}$ & $ \frac{1}{\sqrt{2}}$ \\
  & $(-,0)$ & $-1$ & $ \frac{1}{\sqrt{2}}$ & $ \frac{1}{\sqrt{2}}$ \\
b & $(0,-)$ & $-1$ & $ \frac{1}{\sqrt{2}}$ & $ - \frac{1}{\sqrt{2}}$ \\
  & $(0,0)$ & $-\frac{1}{\sqrt{2}}$ & $ \frac{1}{\sqrt{2}}$ & $ - \frac{1}{\sqrt{2}}$ \\
  & $(+,-)$ & $\frac{1}{\sqrt{2}}$ & $ \frac{1}{\sqrt{2}}$ & $ \frac{1}{\sqrt{2}}$ \\
  & $(+,0)$ & $-1$ & $ \frac{1}{\sqrt{2}}$ & $ \frac{1}{\sqrt{2}}$ \\
\hline
\hline
    \end{tabular}
\end{table}


\subsection{Cross sections}

\label{Cross Sections}

We define the isospin-spin-averaged cross section in the center of mass (CM) frame for the processes in Eq. (\ref{ampl})  as
\begin{eqnarray}
  \sigma_r ^{\left(\varphi \right)}(s) 
= \frac{1}{64 \pi^2 s }  \frac{|\vec{p}_{f}|}{|\vec{p}_i|}  \int d \Omega 
\overline{\sum_{S, I}} 
|\mathcal{M}_r  ^{\left(\varphi \right)} (s,\theta)|^2 ,
\label{eq:CrossSection}
\end{eqnarray}
where $r = a, b$ designates the reactions according to Eq.~(\ref{ampl}); $\sqrt{s}$ is the CM energy;  $|\vec{p}_{i}|$ and $|\vec{p}_{f}|$ denote the three-momenta of initial and final particles in the CM frame, respectively; the symbol $\overline{\sum_{S,I}}$ stands for the sum over the spins and isospins of the particles in the initial and 
final state, weighted by the isospin and spin degeneracy factors of the two particles forming the initial state for the reaction 
$r$, i.e. \cite{XProd1} 
\begin{eqnarray}
\overline{\sum_{S,I}}|\mathcal{M}_r|^2 & \to & 
\frac{1}{g_{1i,r}}
 \frac{1}{g_{2i,r}} \sum_{S,I}|\mathcal{M}_r|^2, 
\label{eq:DegeneracyFactors}
\end{eqnarray}
where $g_{1i,r}=(2I_{1i,r}+1)(2I_{2i,r}+1)$ and $g_{2i,r}= (2S_{1i,r}+1)(2S_{2i,r}+1)$ are the degeneracy factors  of the particles in the initial state, and  
\begin{eqnarray}
\sum_{S,I} |\mathcal{M}_r|^2 = \sum_{Q_{1f}, Q_{2f}} 
\left[\sum_{S}\left|\mathcal{M}_r ^{(Q_{1f},Q_{2f})}\right|^2\right].
\label{eq:Sum}
\end{eqnarray}
Thus, the contributions for the isospin-spin-averaged cross section are distinguished by the possible charge configurations $(Q_{1f}, Q_{2f})$, according to Table~\ref{Tab1}.

Also, we can evaluate the cross sections related to the inverse processes, where
the $X_J$ is produced, using the detailed balance relation.  

Another feature is that when evaluating the cross sections, to prevent the artificial increase of the amplitudes with the energy we make use of a Gaussian form factor defined as~\cite{Huang:2020ptc}: 
\begin{eqnarray}
	F & = & \exp{\left(- \frac{p_E ^2}{\alpha ^2}\right)},
	  \label{formfactor}
\end{eqnarray}
where $p_E$ is the Euclidean Jacobi momentum. The size parameter $\alpha$ is taken according to the choice stated in previous subsection~\footnote{In the context of the method in Ref.~\cite{Huang:2020ptc} (and Refs. therein) based on the Weinberg compositeness condition, the function $F$ is an effective correlation function introduced to reflect the distribution of  two constituents of a bound state system (as in the present case $\bar{D}^*  K^*$), as well as to make the Feynman ultraviolet finite.}.

The computations of the present work are done with the isospin-averaged masses reported in Ref.~\cite{Zyla:2020zbs}: $m_{\pi} = 137.27 \MeV;  m_{K} = 495.65 \MeV;  m_{K^{*}} = 893.61 \MeV; m_{\bar{D}} = 1867.24 \MeV; m_{\bar{D}^*} = 2008.56 \MeV$.



\begin{figure}[!ht]
    \centering
       \includegraphics[{width=8.0cm}]{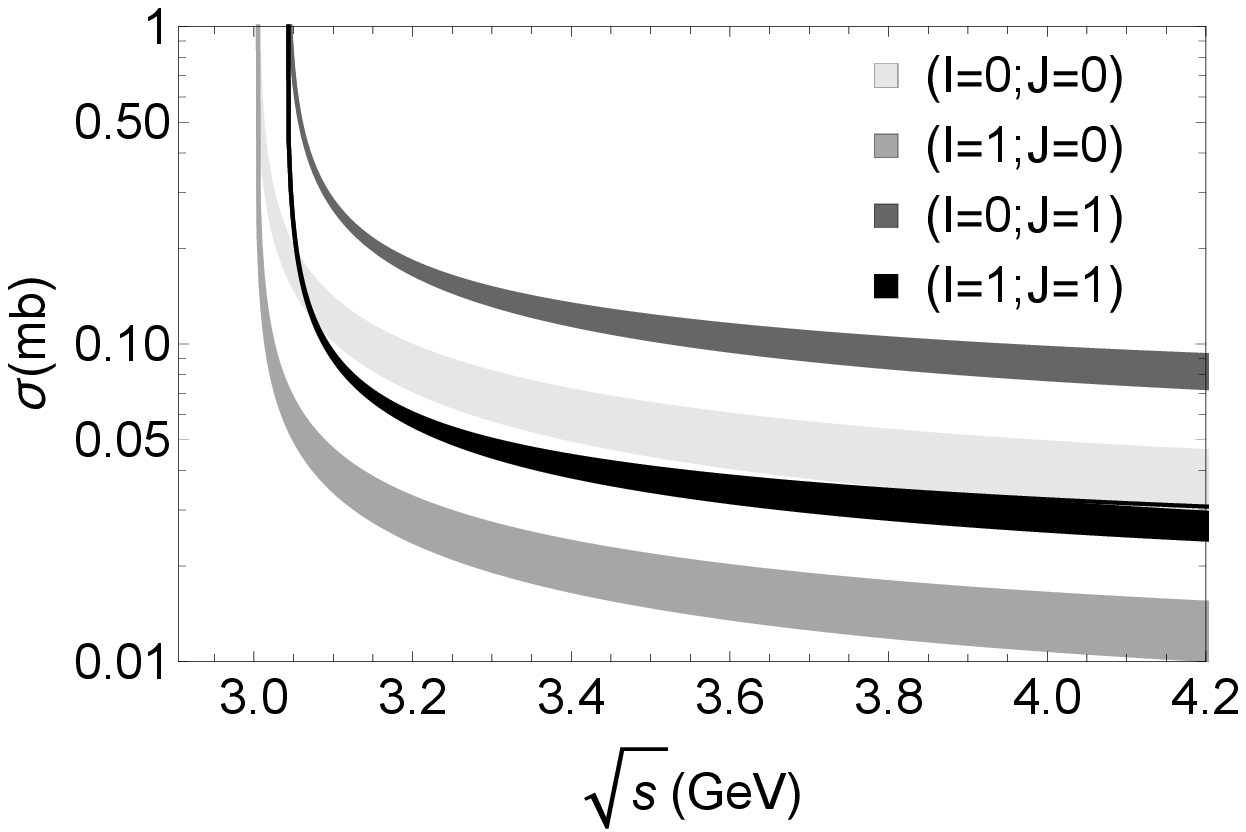} 
       \includegraphics[{width=8.0cm}]{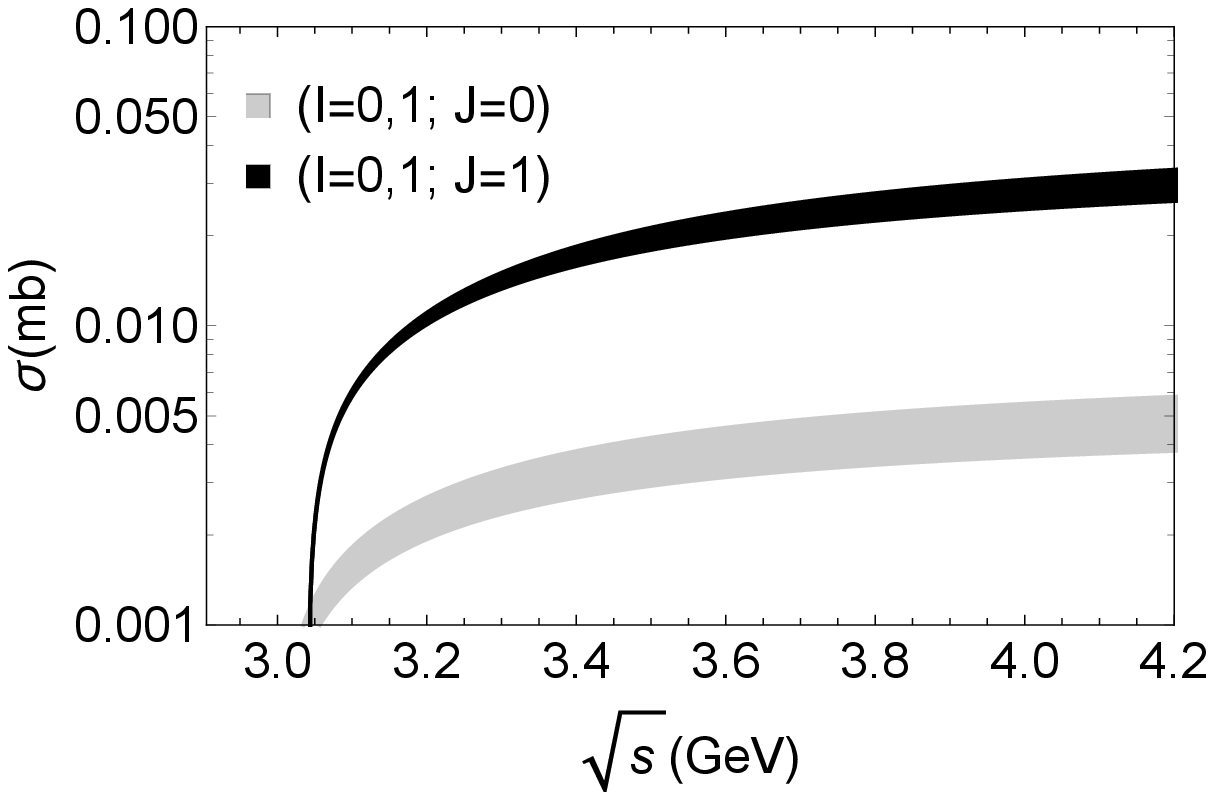} 
        \caption{Top: Cross sections for the absorption process $X_J \pi \to \bar{D}^{\ast} K $, as functions of CM energy $\sqrt{s}$, considering the state $X_J$ with different spin $J$ and isospin $I$. Upper and lower limits of the bands are obtained taking $\alpha$ with lower and upper error limits. Bottom: cross sections for the respective inverse (production) process, obtained via the detailed balance relation. }
    \label{CrSec-DStarK}
\end{figure}

\begin{figure}[!ht]
    \centering
       \includegraphics[{width=8.0cm}]{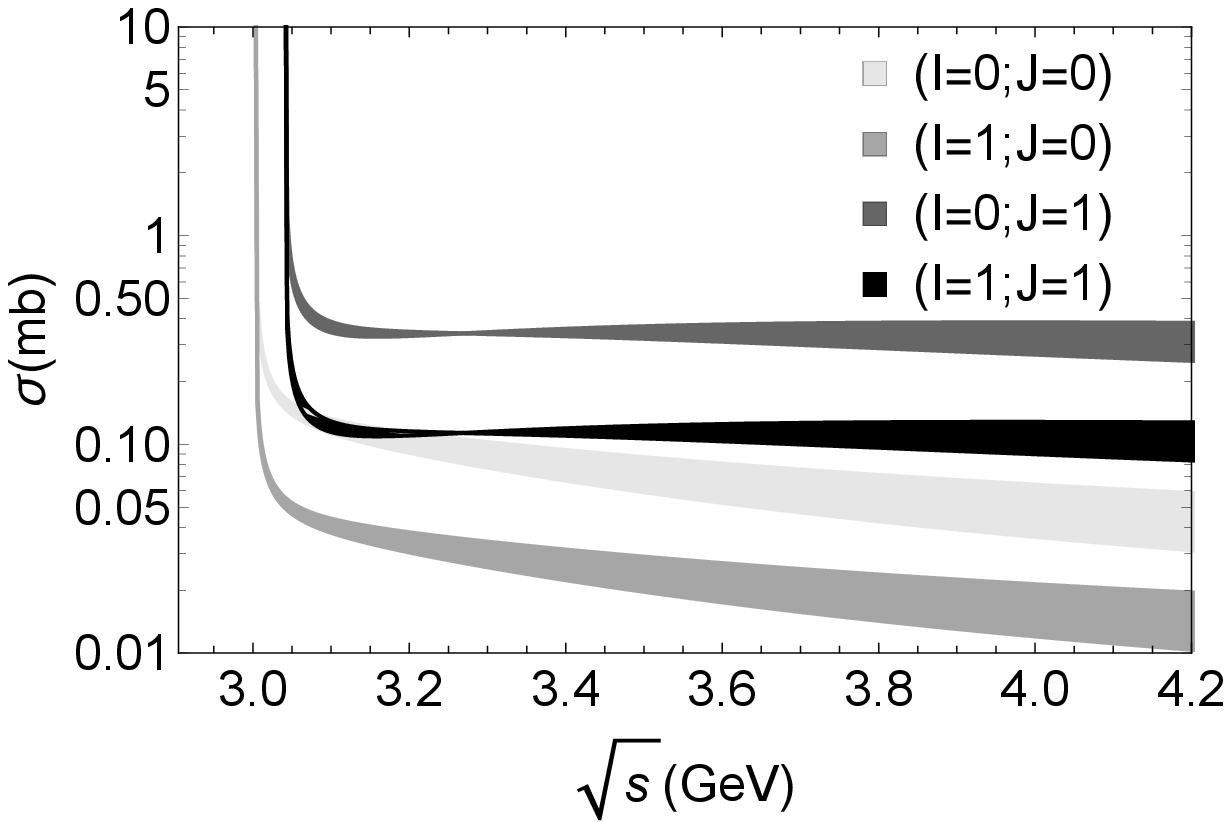} 
       \includegraphics[{width=8.0cm}]{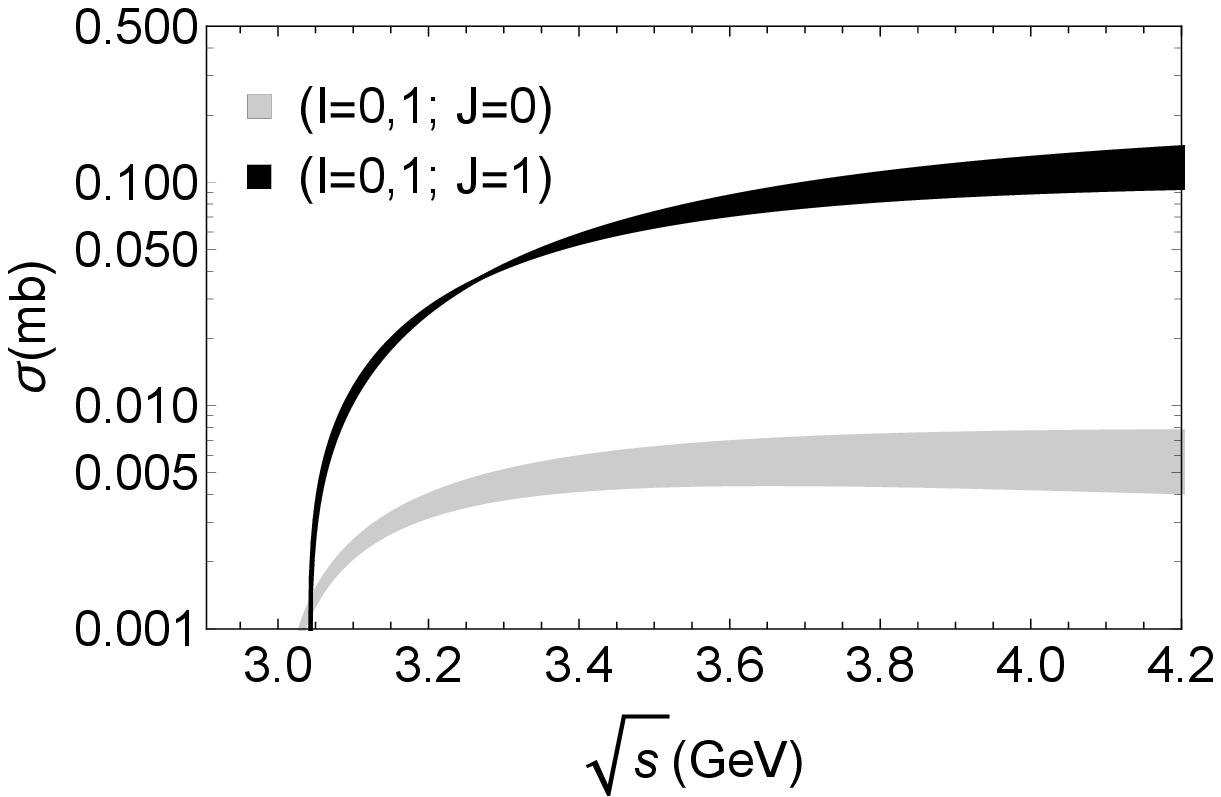} 
        \caption{Top: Cross sections for the absorption process $X_J \pi \to K^{\ast}  \bar{D} $, as functions of CM energy $\sqrt{s}$, considering the state $X_J$ with different spin $J$ and isospin $I$. Upper and lower limits of the bands are obtained taking $\alpha$ with lower and upper error limits. Bottom: cross sections for the respective inverse (production) process, obtained via the detailed balance relation. }
    \label{CrSec-DKStar}
\end{figure}



In Fig.~\ref{CrSec-DStarK} the $X_J \pi$ absorption and production cross sections for the processes involving  the particles $\bar{D}^{\ast}, K $ in the final or initial states are plotted as a function of the CM energy $\sqrt{s}$. 
The absorption cross sections are exothermic and become infinite near
the threshold. Within the range $ 3.2 \leq \sqrt{s} \leq 4.2 \GeV$, they are found to be $ \sim 1 \times 10^{-2} - 1 \times  10^{-1} \mb$, with the magnitudes between the situations involving $X_J$ with same spin but different isospins being distinguishable because of the degeneracy factor. But if we compare the magnitudes of reactions with the same isospin, it can be noticed that the cross sections for $X_1$ are bigger than those for $X_0$ by a factor about $2-3$. In the case of the production cross sections, they are endothermic and there is no distinction between the processes with different isospins. Their magnitudes are $ \sim 2 \times 10^{-3} - 3 \times  10^{-2} \mb$,  within the range $ 3.2 \leq \sqrt{s} \leq 4.2 \GeV$, 
suggesting that the production cross sections are smaller than the absorption ones due to kinematic effects.

The cross sections for the processes involving the particles $K^{\ast}, \bar{D}$ in the final or initial states are plotted in Fig.~\ref{CrSec-DKStar}. In general, the results are similar to the processes analyzed previously, except to the fact that the difference between the cross sections for $X_1$ production become bigger than those for $X_0$ by a factor about $20-25$. Thus, we can infer that the spin-1 state can be formed and absorbed by light mesons more easily than the spin-0 state.

\subsection{Cross sections averaged over the thermal distribution  }

\label{ThermalAvCrossSection} 

Now use the results reported above to compute the thermally averaged cross sections for the $X_J$ production and absorption reactions. To this end, let us introduce the cross section averaged over the thermal distribution for a reaction involving an initial two-particle state going into two final 
particles $ab \to cd$. It is given by~\cite{ChoLee1,XProd2,Koch}
\ben
\langle \sigma_{a b \rightarrow c d } v_{a b}\rangle &  = & 
\frac{ \int  d^{3} \mathbf{p}_a  d^{3}
\mathbf{p}_b f_a(\mathbf{p}_a) f_b(\mathbf{p}_b) \sigma_{a b \rightarrow c d } 
\,\,v_{a b} }{ \int d^{3} \mathbf{p}_a  
d^{3} \mathbf{p}_b f_a(\mathbf{p}_a) f_b(\mathbf{p}_b) }
\nonumber \\
& = & \frac{1}{4 \beta_a ^2 K_{2}(\beta_a) \beta_b ^2 K_{2}(\beta_b) } 
\nonumber \\
& & \times \int _{z_0} ^{\infty } dz  K_{1}(z) \,\,\sigma (s=z^2 T^2) 
\nonumber \\
& & \times \left[ z^2 - (\beta_a + \beta_b)^2 \right]
\left[ z^2 - (\beta_a - \beta_b)^2 \right],
\nonumber \\
\label{thermavcs}
\een
where $v_{ab}$ represents the relative velocity of the  two initial  interacting 
particles $a$ and $b$; the function $f_i(\mathbf{p}_i)$ is the Bose-Einstein   
distribution of particles of species $i$, which depends on the temperature    
$T$; $\beta _i = m_i / T$, $z_0 = max(\beta_a + \beta_b,\beta_c 
+ \beta_d)$, and $K_1$ and $K_2$ the modified Bessel functions of second kind.


\begin{figure}[!ht]
    \centering
       \includegraphics[{width=8.0cm}]{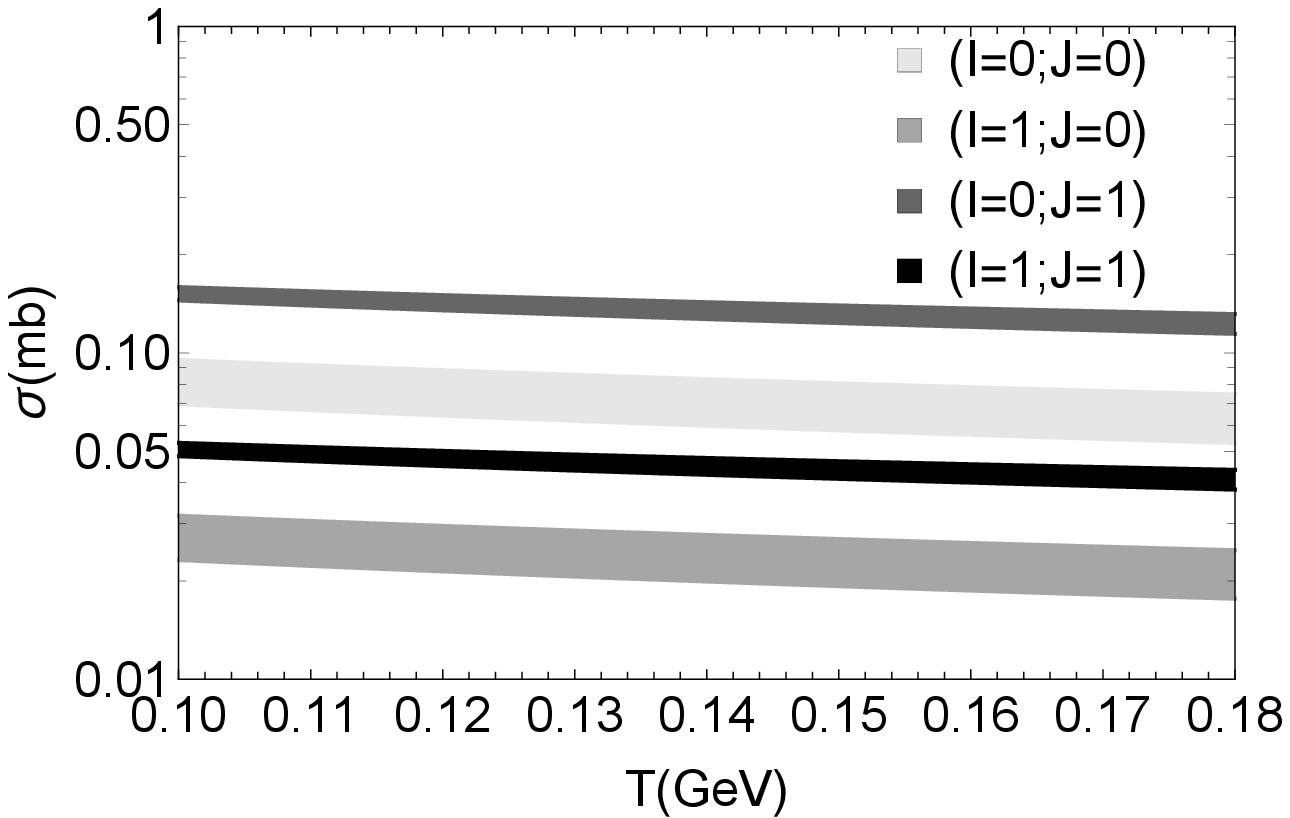} 
       \includegraphics[{width=8.0cm}]{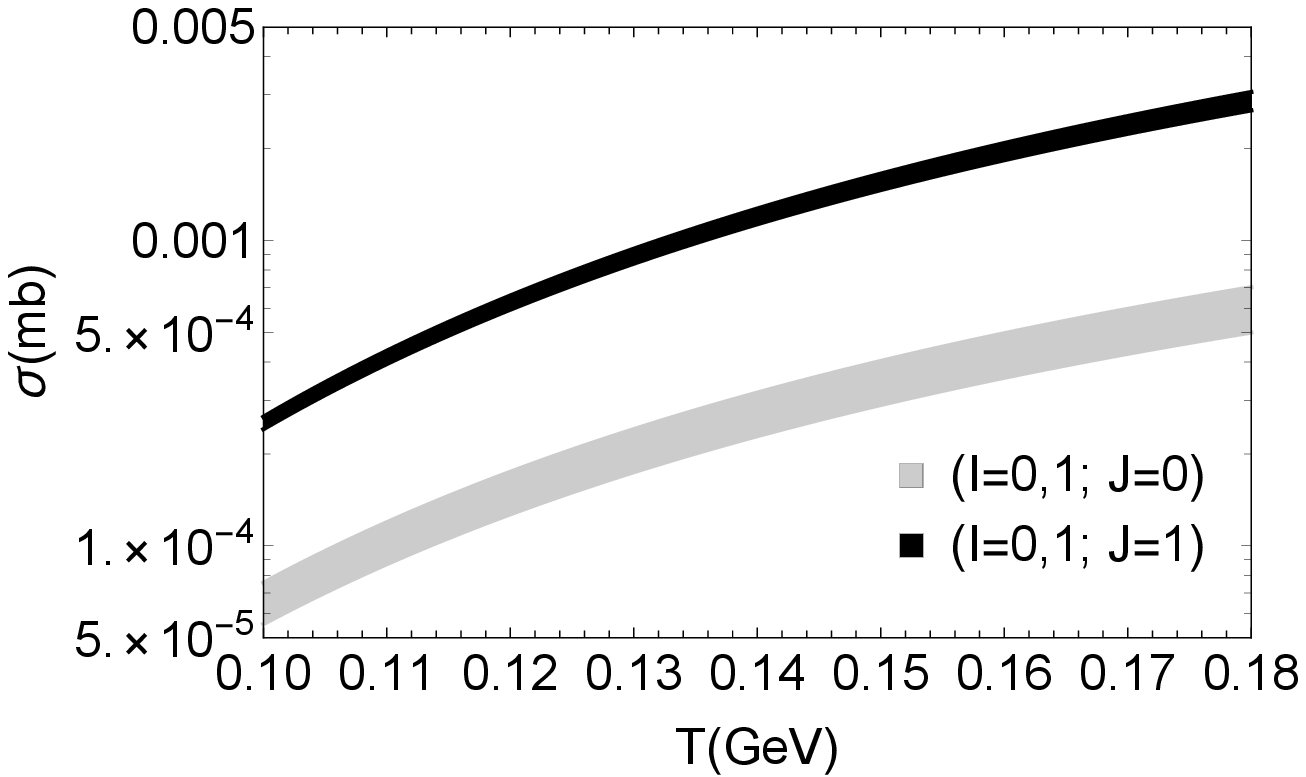} 
        \caption{Top: thermally averaged cross sections for the absorption process $X_J \pi \to \bar{D}^{\ast} K $, as functions of temperature $T$, considering the state $X_J$ with different spin $J$ and isospin $I$.  Upper and lower limits of the bands are obtained taking $\alpha$ with lower and upper error limits. Bottom: thermally averaged cross sections for the respective inverse (production) process. }
    \label{Averaged-CrSec-DStarK}
\end{figure}

\begin{figure}[!ht]
    \centering
       \includegraphics[{width=8.0cm}]{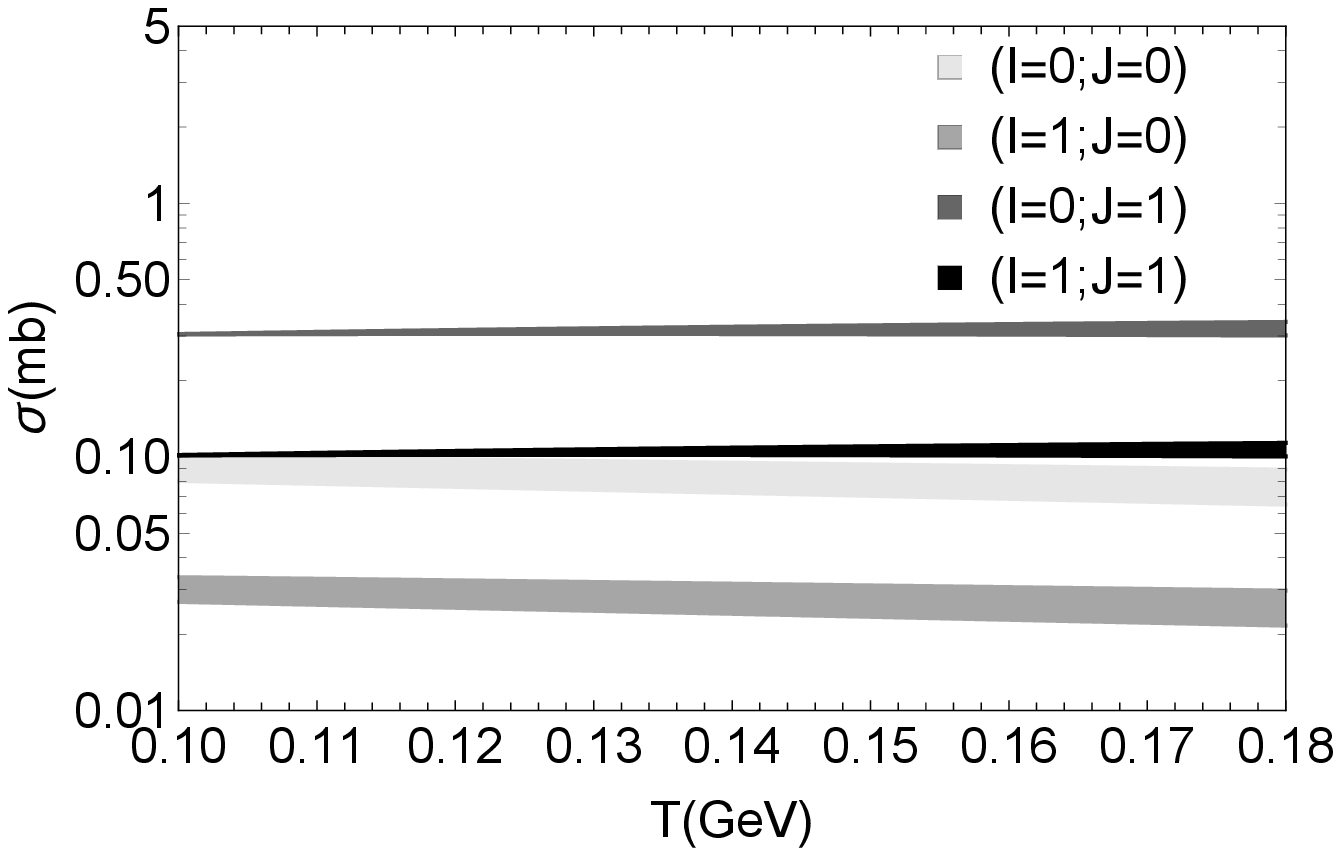} 
       \includegraphics[{width=8.0cm}]{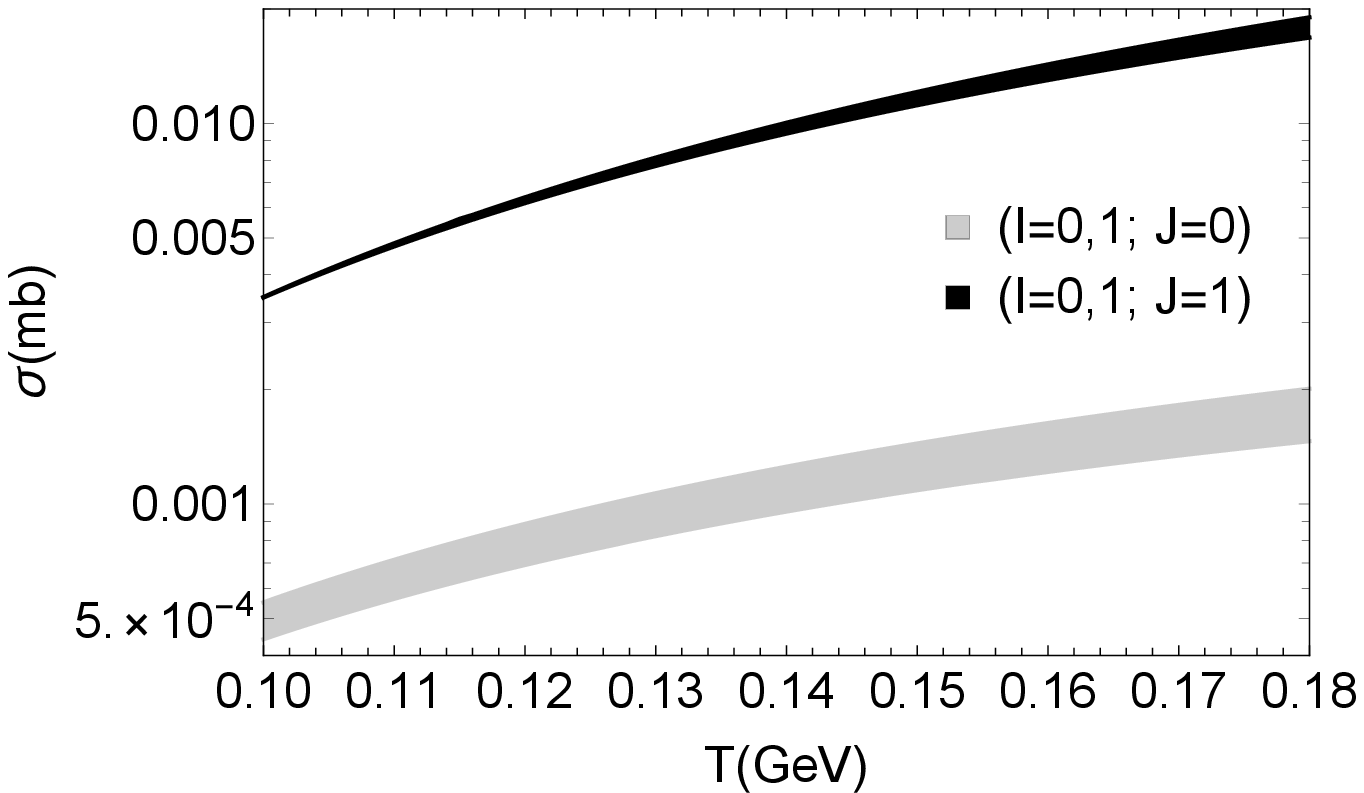} 
        \caption{Top: thermally averaged cross sections for the absorption process $X_J \pi \to K^{\ast}  \bar{D} $, as functions of temperature $T$, considering the state $X_J$ with different spin $J$ and isospin $I$.  Upper and lower limits of the bands are obtained taking $\alpha$ with lower and upper error limits. Bottom: thermally averaged cross sections for the respective inverse (production) process. }
    \label{Averaged-CrSec-DKStar}
\end{figure}


The thermally averaged cross sections for  $X_J \pi $ absorption and production via the processes discussed in the previous subsection are shown in Figs.~\ref{Averaged-CrSec-DStarK} and ~\ref{Averaged-CrSec-DKStar}. The $X_0 \pi$ absorption reactions for final states  $ \bar{D}^{\ast} K $ and $ K^{\ast}  \bar{D} $ have similar magnitudes, almost constant in the considered range of temperature. On the other hand, the thermally averaged cross sections for the production reactions with final state $ K^{\ast}  \bar{D} $ are bigger than the ones with final state $ \bar{D}^{\ast} K$ by a factor about one order of magnitude, and with a relevant change with the temperature. Again, the case with $J=1$ have greater values of $ \langle \sigma_{X_J \pi \to \bar{D}^{\ast} K, K^{\ast}  \bar{D} } v_{a b}\rangle$ with respect to $J=0$ by a factor  about $2-3$, while for the production reaction this factor is about 2 for final state $ \bar{D}^{\ast}  K $ and 10 for final state $ K^{\ast}  \bar{D}$. 

The main point here is that the thermally averaged cross sections for  $X_J \pi$ annihilation and production reactions have different magnitudes, and this might play an important role in the time evolution of the  $X_J$ multiplicity, 
 to be analyzed in the next section.

\section{Time evolution of  $X_J$ abundance }

\label{abundance}

Now we study the time evolution  of the $X_J$ abundance in hadronic matter, using the thermally averaged cross sections estimated in the previous  section. We focus on the influence of $ X_J - \pi $  interactions on the abundance of  $X_J$ during the hadronic stage of heavy ion collisions. The momentum-integrated evolution equation for the abundances of particles included in the processes previously discussed reads~\cite{Chen:2007zp,ChoLee1,XProd2,Koch,Cho:2015qca,UFBaUSP1,Abreu:2017cof,Abreu:2018mnc}
\begin{eqnarray} 
\frac{ d N_{X_J} (\tau)}{d \tau} & = & \sum_{ a b } 
\left[ \langle \sigma_{ a b  \rightarrow X_J \pi} 
v_{ a b} \rangle n_{a} (\tau) N_{b}(\tau)
 \right. \nonumber \\ & &  \left. 
- \langle \sigma_{ X_J \pi \rightarrow a b  } v_{X_J \pi } 
\rangle n_{\pi} (\tau) N_{X_J}(\tau) 
\right] \nonumber \\  
& & + \left. \langle \sigma_{ \bar{D} K \rightarrow X_J} 
v_{  \bar{D} K } \rangle n_{ \bar{D} } (\tau) N_{K}(\tau)
 \right. \nonumber \\ 
& &  -  \langle \Gamma_{X_J \to \bar{D} K} \rangle N_{X_J} (\tau)  , 
\label{rateeq}
\end{eqnarray}
where $n_{i} (\tau)$ are $N_{i}(\tau)$ denote  the density and          
the abundances of involved mesons in hadronic matter at  
proper time $\tau$. 
Since the lifetime of the $X_J$ states is less than that of the hadronic stage presumed in this 
work (of the order of 10 fm/c), thus the decay of $X_J$ and its regeneration from the 
daughter particles  $ \bar{D} $ and $K$ are included in the last two lines of the rate equation. We adopt a similar approach to Ref.~\cite{Cho:2015qca}: the scattering cross section for the $X_J$ state production from  $ \bar{D} $ and $K$ mesons is given by the spin-averaged relativistic Breit-Wigner cross section, 
\begin{eqnarray} 
  \sigma _{\bar{D} K \to X_J} = \frac{g_{X_J}}{g_{\bar{D} } g_{K}} \frac{4 \pi}{p_{cm} ^2 }  \frac{s \Gamma_{X_J \to \bar{D} K} ^2}{\left( s - m_{X_J}^2 \right)^2 + s  \Gamma_{X_J \to \bar{D} K} ^2 }, 
  \label{CrSecFormX}
 \end{eqnarray}
where $g_{X_J}, g_{\bar{D} } $ and $g_{K}$ are the degeneracy of $X_J$, $\bar{D}$ and $K$ mesons, respectively; $p_{cm}$ is the momentum in CM frame; $\Gamma_{X_J \to \bar{D} K}$ is the total decay width for the reaction $X_J \to \bar{D} K$, which is supposed to be effectively $\sqrt{s}$-dependent in the form 
\begin{eqnarray} 
  \Gamma_{X_J \to \bar{D} K} (\sqrt{s}) = \frac{1}{32 \pi^2 s }   |p_{cm}(\sqrt{s})| \overline{\sum} 
|\mathcal{M}_{\Gamma}  |^2   d \Omega 
    \label{decayX}
\end{eqnarray}
with $ \mathcal{M}_{\Gamma} \propto g_{X_J\bar{D} K} \{ 1, \epsilon (p_X) \cdot (p_{\bar{D}} - p_K)\}$ for $J=0$ and $J=1$, respectively; the value of constant $g_{X_J\bar{D} K}$ is determined from the experimental value of $ \Gamma_{X_J \to \bar{D} K} (\sqrt{s}) $ with the system in the rest frame of the $X_J$: $\Gamma _{X_0 } = 57.2 \MeV , \Gamma _{ X_1} =  110.3 \MeV $. It can be noticed that the cross section for the $X_J$ state production in Eq.~(\ref{CrSecFormX}) are about 2.5-8 mbarn for $J=0$, and about 6.5-20 mbarn for $J=1$, depending on the isospin. Its average  over the thermal distribution, $\langle \sigma_{ \bar{D} K \rightarrow X_J} v_{  \bar{D} K } \rangle$, can be determined from Eq.~(\ref{thermavcs}). The thermally averaged decay width of $X_J$ is given by~\cite{Cho:2015qca}
 \begin{eqnarray} 
 \langle \Gamma_{X_J \to \bar{D} K} \rangle =  \Gamma_{X_J \to \bar{D} K} \left(m_{X_J} \right) \frac{ K_1  \left(m_{X_J} / T \right) }{ K_2  \left(m_{X_J} / T \right)} . 
    \label{TherhmaldecayX}
\end{eqnarray}

In the obtention of solutions of Eq.~(\ref{rateeq}), we assume that the pions, charm  and strange mesons in the reactions contributing to the $X_J $ abundance are in equilibrium. Therefore, the density $n_{i} (\tau)$ can be 
written as~\cite{ChoLee1,XProd2,Koch,Cho:2017dcy}
\ben n_{i} (\tau) &  \approx & \frac{1}{2 \pi^2}\gamma_{i} g_{i} m_{i}^2 
T(\tau)K_{2}\left(\frac{m_{i} }{T(\tau)}\right), 
\label{densities}
\een
where $\gamma _i$ and $g_i$ are the fugacity factor and the  degeneracy factor of the particle, respectively. In this sense, the multiplicity $N_i (\tau)$ is obtained by multiplying the density $n_i(\tau)$ by the volume $V(\tau)$. 
The time dependence of $n_{i} (\tau)$ is inserted through the parametrization of the temperature $T(\tau)$ and volume $V(\tau)$ used to model the dynamics of relativistic heavy ion collisions after the end of the QGP phase. The 
hydrodynamical expansion and cooling of the hadron gas are based on the boost invariant Bjorken picture with an accelerated transverse expansion~\cite{Cho:2017dcy,Cho:2015qca,ChoLee1}. Accordingly,  the $\tau$ dependence of $V(\tau)$ and $T$ are given by
\ben
T(\tau) & = & T_C - \left( T_H - T_F \right) \left( \frac{\tau - \tau _H }
{\tau _F -  \tau _H}\right)^{\frac{4}{5}} , \nonumber \\V(\tau) & = & \pi \left[ R_C + v_C 
\left(\tau - \tau_C\right) + \frac{a_C}{2} \left(\tau - \tau_C\right)^2 
\right]^2 \tau \, c , \nonumber \\
\label{TempVol}
\een
where $R_C $ and $\tau_C$  label the final transverse  and longitudinal sizes of the QGP; $v_C $ and  $a_C $ are its transverse flow velocity and transverse  acceleration at $\tau_C $; $T_C$ is the critical temperature for the QGP to hadronic matter transition; $T_H $  is the temperature of the hadronic matter at the end of the mixed phase, occurring at the time $\tau_H $; and the kinetic freeze-out temperature  $T_F $ leads to a freeze-out time $\tau _F $~\cite{Cho:2017dcy,Cho:2015qca,ChoLee1}. We remark that the Bjorken picture is an attempt to describe in a simple way the hydrodynamic evolution of fluid, and its improvement is a project beyond the scope of this work. Thus, we employ this parametrization as an exploratory tool that properly characterizes the essential features of hydrodynamic expansion and cooling of the hadron gas for our purposes.

The evolution of $X_J$ multiplicity is evaluated with the hadron gas formed in central $Pb-Pb$ collisions at $\sqrt{s_{NN}} = 5$ TeV at the LHC. Keeping in mind that there is no unique choice of parameters used as input for Eq. (\ref{TempVol}) in literature, our option has been based on the set listed in Table 3.1 of Ref.~\cite{Cho:2017dcy}, summarized in Table~\ref{param} for completeness. 
In this sense, we obtain the values for the volume of the system
$V_H \equiv V(\tau_H)$ and $V_F \equiv V(\tau_F) $ in reasonable agreement with those in~\cite{Cho:2017dcy}. 
In the case of the cooling function $T(\tau)$, this parametrization gives also a similar behavior to others works (see for example~\cite{Abreu:2018mnc,Hong:2018mpk}).

\begin{center}
\begin{table}[h!]
\caption{Parameters used in the parametrization of the hydrodynamical expansion in the scenario of central $Pb-Pb$ collisions at $\sqrt{s_{NN}} = 5$ TeV at the LHC~\cite{Cho:2017dcy}, given by Eq.~(\ref{TempVol}). }
\vskip1.5mm
\label{param}
\begin{tabular}{ c c c }
\hline
\hline
 $v_C$ (c) & $a_C$ (c$^2$/fm) & $R_C$ (fm)   \\   
0.5 & 0.09 & 11  
\\  
\hline
 $\tau_C$ (fm/c) & $\tau_H$ (fm/c)  &  $\tau_F$ (fm/c)  \\   
7.1  & 10.2 & 21.5
\\  
\hline
  $T_C (\MeV)$  & $T_H (\MeV)$ & $T_F (\MeV)$ \\   
 156 & 156 & 115   \\  
\hline
 $N_c$  & $N_{\pi}(\tau_F)$ & $N_{K}(\tau_H)$ \\   
 14 & 2410 & 134 \\  
\hline
\hline
\end{tabular}
\end{table}
\end{center}

We also suppose that the total number of charm quarks, $N_c$,  in charm hadrons is  conserved during the processes, i .e. $n_c(\tau) \times V(\tau) = N_c$. This implies that the charm quark fugacity factor $\gamma _c $ in Eq.~(\ref{densities})  is time-dependent in order to keep $N_c$ constant. The total numbers of  pions and strange mesons at freeze-out were also based on 
Ref.~\cite{Cho:2017dcy}. Considering that the pions and strange mesons might be out of chemical equilibrium in the later part of the hadronic evolution, they also have time dependent fugacities.


%
%
%
%
%
%
%
%
%
%
%
%
%
%
%
%
%
%
%


We study the evolution of yields obtained for the $X_J $   
abundance in different approaches: the statistical and the 
coalescence models. In the statistical model, hadrons are produced in 
thermal and chemical equilibrium, according to the expression of the 
abundance obtained from Eq.~(\ref{densities}) of the corresponding hadron. 
%
%
%
%
It can be noticed that the  information 
associated  to the internal structure of state is not taken into account.        
In the coalescence model, however, the determination of the yield of a hadron is related on the overlap 
of the density matrix of the constituents in an emission source with 
the Wigner function of the produced particle. The information on the internal structure of the considered hadron, such as 
angular momentum, multiplicity of quarks, etc., is contemplated.                           
Then, following Refs.~\cite{Cho:2017dcy,Cho:2015qca,ChoLee1}, the number 
of $X_J$'s produced  is given by:
\begin{eqnarray}
  N_{X_{J}} ^{Coal} & \approx & g_{X_{J}} \prod _{j=1} ^{n} \frac{N_j}{g_j} 
\prod  _{i=1} ^{n-1} 
\frac{(4 \pi \sigma_i ^2)^{3/2} }{V (1 + 2 \mu _i T \sigma _i ^2 )} 
\frac{(2 l_i)!! }{V (2 l_i+1)!!}  \nonumber \\
& &  \times   \left[ \frac{4 \mu_i T \sigma_i ^2 }{ (1 + 2 \mu _i T \sigma _i ^2 ) }
\right]^{l_i}
\nonumber \\
 & \approx & \frac{g_{X_{J}}(M\omega)^{3/2}}{(4\pi)^{3/2}}  \frac{(2 T / \omega )^L }{ (1 + 2T / \omega)^{n+L-1} }
\nonumber \\
& & \times   \prod _{j=1} ^{n} \frac{N_j (4\pi)^{3/2}}{g_j (m_j \omega)^{3/2}} 
\prod  _{i=1} ^{n-1} 
\frac{(2 l_i)!! }{V (2 l_i+1)!!} , 
\label{NXCoal}
\end{eqnarray}
where $g_j$ and $N_j$ are the degeneracy and number of the $j$-th constituent of the $X_J$; $L= \sum_{i=1} ^{n-1} l_i $; $M = \sum_{i=1} ^{n} m_i $;  $\sigma _i = (\mu _i \omega)^{-1/2}$ is related to the reduced constituent masses $\mu _i=m_{i+1} ^{-1}+ \left(\sum_{j=1} ^{i} m_j \right)^{-1}$ and the oscillator
frequency $\omega $, assuming an harmonic oscillator prescription for the hadron internal structure; and $l_i$ is the angular momentum of the internal (relative) wave function associated with the relative coordinate $y_i$, being 0 for an $S$-wave, 1 for a $P$-wave constituent, and so on. The relation $ \mu_i \sigma _i ^2 =  \omega ^{-1}$   was used to convert the dependence on $l_i$ into the form of the orbital angular momentum sum $L$, which makes the last factor in last line of Eq.~(\ref{NXCoal}) depend on the way that $L$ is decomposed into $l_i$ if $L \geq 2$. Then, let us first consider $X_J$ as a tetraquark state, produced via quark coalescence mechanism from  the QGP phase at the critical temperature $T_c$ when the volume is $V_C$; in this situation, $n=3$ and $L=l_1+l_2$. Remarking that the $X_0 \rightarrow D^- K^+ $ decay is $S$-wave, while the $X_1 \rightarrow D^- K^+ $ decay is $P$-wave, then for $J=0$ we have the combination $(l_1,l_2)=(0,0)$ (which makes the mentioned factor equal to 1), and for $J=1$ the possibilities $(l_1,l_2)=(0,1)$ or $(1,0)$ (which gives the factor 2/3). On the other hand, for the yields of $X_J$ as a weakly bound hadronic molecule from the coalescence of hadrons, they are evaluated at the kinetic freeze-out temperature $T_F$ and volume $V_F$. On this  regard, $J=0(1)$ should be a $S(P)$-wave hadronic molecule, as speculated in Ref.~\cite{Huang:2020ptc}. In Table~\ref{Tab3} we give the yields in these different approaches. The comparison between the values of $  N_{X_{J}}  ^{0 (Stat.)}  $ and $ N_{X_{J}}  ^{0  (4q)}$ indicates that the number of $X_J $'s calculated with the statistical model is   
greater than the  four-quark state (formed by quark coalescence)  by about one order of magnitude.  Also, for tetraquark coalescence the case with $J=1$ has a smaller initial yield than the state with $J=0$.

In the present context we evaluate the time evolution of the $X_J$ abundance by solving Eq.~(\ref{rateeq}), with initial conditions given within statistical and tetraquark coalescence models. We emphasize that in the case of molecular states,  since they are dominantly formed by hadron coalescence at the end of the hadronic phase, we use the yields shown in last column of Table~\ref{Tab3} just for comparison with the results obtained from the time evolution of initial yields in statistical and tetraquark coalescence models up to kinetic freeze-out, i.e. $N_{X_J} (\tau _F)$.

\begin{center}
\begin{table}[h!]
\caption{The $X_J$ yields in central $Pb-Pb$ collisions at $\sqrt{s_{NN}} = 5$ TeV at the LHC using statistical and molecular/four-quark coalescence models, according to Eqs.~(\ref{densities}) and (\ref{NXCoal}). The oscillator frequency for tetraquark states produced via quark coalescence mechanism  is $\omega = 588 \MeV$~\cite{Chen:2007zp,Cho:2017dcy}. In the case of molecular states formed by hadron coalescence at the end of the hadronic phase, we have used  $\omega = 6 B = 217 \MeV$ and $ 108 \MeV$ for a $S$-wave and a $P$-wave, respectively ($B$ being the binding energy).  }
\label{Tab3}
\begin{tabular}{c | c c c }
\hline
\hline
State & $  N_{X_{J}}  ^{ (Stat.)} (\tau_H) $  & $  N_{X_{J}}  ^{  (4q-Coal.)} (\tau_C) $ & $ N_{X_{J}}   ^{0  (Mol.-Coal.)}  (\tau_F) $    \\   
\hline
$J=0, I=0$ & $ 6.9 \times 10^{-3}$ & $ 1.3 \times 10^{-3}$ & $ 4.5 \times 10^{-4}$
\\  
$J=0, I=1$ &  $ 2.1 \times 10^{-2}$ & $ 3.9 \times 10^{-3}$ & $ 1.36 \times 10^{-3}$
\\  
\hline
$J=1, I=0$ &  $ 1.7 \times 10^{-2}$ & $ 9.0 \times 10^{-4}$ & $ 7.2 \times 10^{-3}$
\\  
$J=1, I=1$ &  $ 5.0 \times 10^{-2}$ & $ 2.7 \times 10^{-3}$ & $ 2.2 \times 10^{-2}$
\\  
\hline
\hline
\end{tabular}
\end{table}
\end{center}


\begin{figure}[!ht]
\includegraphics[{width=8.0cm}]{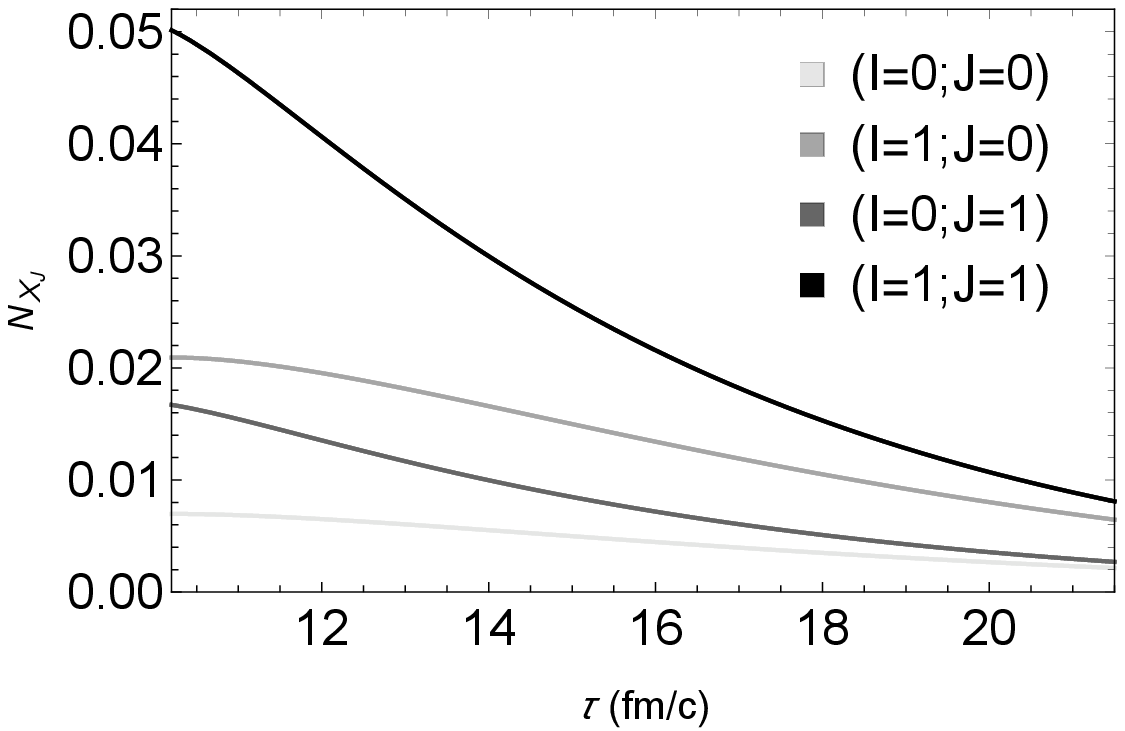}\\ 
\includegraphics[{width=8.0cm}]{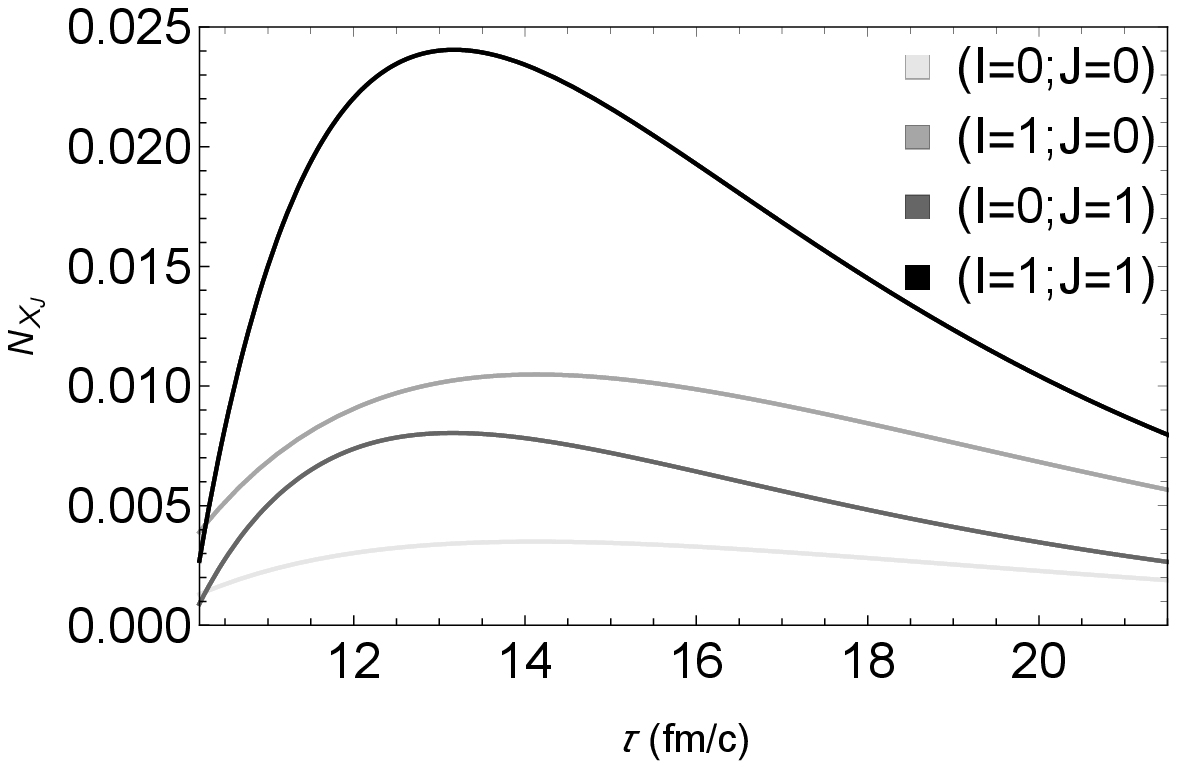}
\caption{Time evolution of the $X_J$ abundance as a function of 
the proper time in central $Pb-Pb$ collisions at $\sqrt{s_{NN}} = 5$ TeV at the LHC.    
Upper panel: Shaded bands represent the evolution of the number of  
$X_J$'s produced  at the end of the mixed phase 
calculated using statistical model for different values of $J,I$. Lower panel: 
Same  as in upper panel but using four-quark coalescence model.}
\label{TimeEvolXJ}
\end{figure}
%

\begin{figure}[!ht]
\includegraphics[{width=8.0cm}]{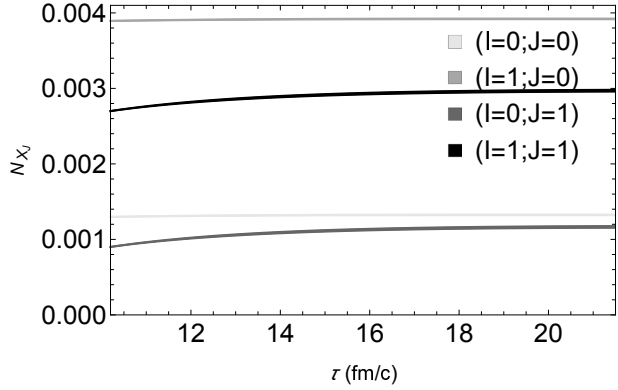}
\caption{Time evolution of the $X_J$ abundance as a function of the proper time in central $Pb-Pb$ collisions at $\sqrt{s_{NN}} = 5$ TeV at the LHC, with the number of $X_J$'s produced  at the end of the mixed phase calculated using four-quark coalescence model, and taking $\Gamma _{X_0, X_1} = 0 \MeV$ in the terms of the last two lines of Eq.~(\ref{rateeq}). }
\label{TimeEvolXJGammaZero}
\end{figure}

In Fig.~\ref{TimeEvolXJ} we show the time evolution of the $X_J$ abundance as a  
function of the proper time, using $ N_{X_{J}}  ^{ (Stat.)}  $ and 
$ N_{X_{J}}  ^{  (4q-Coal.)}$ reported in Table~\ref{Tab3} as initial conditions. The findings suggest that the interactions between the $X_J$'s and the pions through the reactions discussed previously during the hadronic stage of HICs produce relevant changes. Besides, it can be seen that the results are strongly dependent of the initial yields. With the initial conditions calculated from  statistical model, the abundances suffer a reduction  which is more pronounced for the state with $J=1$. On the other hand, when we use the initial conditions at the end of the mixed phase from four-quark coalescence model, we see that the $X_J$ multiplicities experiment an increasing by a factor about 1.5 for $J=0$ and 3 for $J=1$. It is also worth mentioning that the modifications in the abundances are mostly due to the terms associated to the spontaneous decay/regeneration of $X_J$ (i.e. the terms in the last two lines of Eq.~(\ref{rateeq})). To illustrate this issue, in Fig.~(\ref{TimeEvolXJGammaZero}) we plot the time evolution of the $X_J$ abundance using four-quark coalescence model to fix the initial conditions, but taking $\Gamma _{X_0, X_1} = 0 \MeV$. Within these conditions, there is a relative equilibrium between production  and absorption reactions, resulting in a number of  $X_J$'s  throughout the hadron gas phase nearly constant in the case $J=0$; for $J=1 $ is suggested an increasing of the multiplicity  by a factor  about $25\% $ and $10\%$ in the cases of $I=0$ and $I=1$, respectively.

\begin{figure}[!ht]
\includegraphics[{width=8.0cm}]{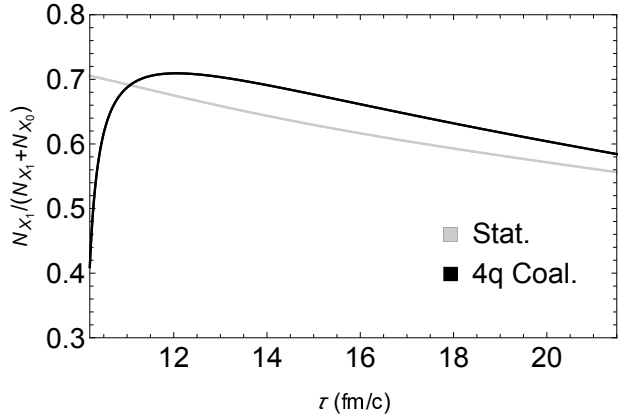}
\caption{Time evolution of the ratio of the $X_1$ abundance to the sum of the  $X_0$ and $X_1$ abundances, as a function of 
the proper time in central $Pb-Pb$ collisions at $\sqrt{s_{NN}} = 5$ TeV at the LHC.}
\label{TimeEvolXJJ0OVERJ0PLUSJ1}
\end{figure}

We also show in Fig.~\ref{TimeEvolXJJ0OVERJ0PLUSJ1} the evolution of the ratio of the $X_1$ abundance to the sum of the  $X_0$ and $X_1$ abundances. In the context of initial conditions with the statistical model, the ratio monotonically decreases from $70\%$ to $56\%$. Using the tetraquark coalescence model, however, the ratio undergoes an abrupt growth from $41\%$ to $71\%$, and a further reduction up to $58\%$ in the end.

%

Finally, we stress that the outputs reported above are based on the evolution of the $X_J$ multiplicities which come from the quark-gluon plasma and might be modified due to hadronic effects via the considered processes in previous sections. In this scenario,  with the initial yields calculated in tetraquark coalescence model ($X_0$ and $X_1$ as $S$-wave and $P$-wave tetraquark states), we have obtained estimations for the case of reminiscent $X_J$ abundances. As stated before, another possible formation  mechanism of $X_J$ is the hadron coalescence, dominant at the end of the hadronic phase.  According to Table~\ref{Tab3} and the results from Fig.~\ref{TimeEvolXJGammaZero} (taking $\Gamma _{X_0, X_1} = 0 \MeV$), the yield of hadronic molecular state for $J=0$ is about 3 times smaller and for $J=1$ is 6 times greater than the contribution from the tetraquark state at kinetic freeze-out. When the spontaneous decay/regeneration of $X_J$ is taken into account (Fig.~\ref{TimeEvolXJ}), these ratios are about 4 times smaller for $J=0$  and 2.5 greater for $J=1$. Hence, remarking that the calculated cross sections does not account for the size of the hadrons, our findings suggest that at the end of the hadronic phase the production of $X_0$ as a hadronic molecular state is reduced with respect to tetraquark state, while for the case of $X_1$ the most prominent production comes from the hadronic molecular state.

\section{Concluding remarks}

\label{Conclusions}

We have investigated in this work the hadronic effects on the recently observed $X_{J=0,1} (2900) $ states in heavy ion collisions. We have made use of Effective Lagrangians to calculate the cross sections and their thermal averages of the processes $X_J \pi \to \bar{D}^{*} K , K^{*} \bar{D}  $,  as well as those of the corresponding inverse processes. Considering also the possibility of different isospin assignments ($I=0,1$), we have found that the magnitude of the cross sections and their thermal averages depend on the quantum numbers, since the energy dependence is  different for $J=0,1$. 
In general the cross sections for $X_1$ are bigger than those for $X_0$, and those involving $X_J \pi$ production are smaller than the absorption ones due to kinematic effects.  As a consequence the thermally averaged cross sections for  $X_J \pi$ annihilation and production reactions also have different magnitudes.

Taking the thermally averaged cross sections as inputs, we have solved the rate equation 
to determine the time evolution of the  $X_J (2900)$  multiplicities. We have found that the $X_J$ abundance is also strongly affected by the quantum numbers $I,J$ during the expansion of the hadronic matter. Considering the $X_J$ as a tetraquark state produced via quark coalescence mechanism from the QGP phase, in which $X_0$ is a relative $S$-wave and $X_1$ a $P$-wave, then when we neglect the $X_J$ spontaneous decays/regenerations their multiplicities are not significantly affected by the interactions with the pions; and hence the number of $X_J$'s would remain essentially unchanged during the  
hadron gas phase in the case of $J=0$.  But the inclusion of these effects (present in the last two lines of Eq. (\ref{rateeq}) ) leads to an increasing by a factor about 1.5 for $J=0$ and 3 for $J=1$ at kinetic freeze-out. Besides, concerning the evolution of the ratio of the $X_1$ abundance to the sum of the  $X_0$ and $X_1$ abundances, it undergoes a growth from $41\%$ at the end of mixed phase to $58\%$ in the end of hadronic phase.

When we compare the multiplicity of  $X_J (2900)$ as hadronic molecular states ($J=0$ as a $S$-wave and $J=1$ as a $P$-wave) and tetraquark states at kinetic freeze-out, the production of $X_0$ as a hadronic molecular state is smaller than the tetraquark state, while for the case of $X_1$ the most prominent production comes from the hadronic molecular state. therefore, as pointed out in analyses performed in the scenario of other exotic states~\cite{ChoLee1,XProd1,XProd2,UFBaUSP1},  we believe that the evaluation of the $X_J (2900)$ abundance in relativistic heavy ion collisions might shed some light on the discrimination of its structure, although the ratio between these two approaches obtained in the present work is not large as in the case of $X(3872)$ discussed in ~\cite{ChoLee1,XProd2}. In our case, if a vertex detector is able to cumulate a number of charmed mesons by about $10^3$, a few $X_0 (2900)$ and $X_1 (2900)$ are expected to be yielded if they are $S$-wave tetraquark state and $P$-wave hadronic molecular state for $J=0$ and $J=1$, respectively. In this sense, in the near future we intend to refine this phenomenological description working on two fronts. The first one is the inclusion of reactions involving light mesons of other species in initial/final states, although we expect that the processes with pions provide the main contributions due to their large multiplicity with respect to other light hadrons. The second one is the improvement of the parametrization of the hydrodynamic evolution, which in principle will allow us to have a more compelling perspective on the role of the hot hadronic medium in the evolution of the $X_J(2900)$ states yielded at the end of the QGP phase.

\begin{acknowledgements}

The author would like to thank the Brazilian funding agencies for their financial support: CNPq (contracts 308088/2017-4 and 400546/2016-7) and FAPESB (contract INT0007/2016).

\end{acknowledgements} 



\end{document}